\documentclass[11pt]{article}
\usepackage{fullpage, graphicx,amsmath,amssymb,cancel,float,color,mathrsfs,amsfonts,eufrak,amsthm}
\usepackage{bbm}
\usepackage{subfigure}
\usepackage{cite}
\usepackage{enumerate}
\usepackage[affil-it]{authblk}
\usepackage{blindtext}
\usepackage{abstract}

\newtheoremstyle{remboldstyle}
  {}{}{}{}{\bfseries}{.}{.5em}{{\thmname{#1 }}{\thmnumber{#2}}{\thmnote{ (#3)}}}
\theoremstyle{remboldstyle}

\newcommand{\drop}[1]{}

\begin{document}
\title{ Coupled Grain Boundary Motion with Triple Junctions}
\author{Anup Basak and Anurag Gupta\thanks{ag@iitk.ac.in (corresponding author)}}
\date{{\small Department of Mechanical Engineering, Indian Institute of Technology, Kanpur 208016, India
\\  \today}}

%\twocolumn[
%  \begin{@twocolumnfalse}
 \maketitle

%%%%%%%%%%%%%%%%%%%%%%%%%%%%%%%%%%%%%%%%%%%%%%%%%%%%%%%%%%
\begin{abstract}
Coupled grain boundary (GB) motion has been studied in a two-dimensional tricrystal where a cylindrical grain is embedded at the center of the planar GB of a large bicrystal. Kinetic relations for GB dynamics, grain orientations, and junctions have been derived within the framework of Gibbs thermodynamics. These are solved numerically to investigate the shrinkage of the embedded grain while emphasizing the role of coupled motion as well as junction mobility in the shape evolution.  

\vspace{2mm}\noindent {\bf Keywords:} Coupled grain boundary motion; Triple junction; Kinetic relation; Tricrystal; Nanocrystalline material
\end{abstract}
%  \end{@twocolumnfalse}
%]
%\saythanks
%%%%%%%%%%%%%%%%%%%%%%%%%%%%%%%%%%%%%%%%%%%%%%%%%%%%%%%%%

A relative tangential motion of neighboring grains coupled with grain boundary (GB) migration is called coupled GB motion \cite{cahn1}. For a grain embedded within a polycrystal the tangential motion is manifested as the relative rotation of the grain with respect to the adjoining grains. It is accomplished through either a pure viscous sliding or a tangential motion geometrically coupled with GB migration, or a combination of both \cite{cahn1}. Whereas GB migration is the dominant mechanism of grain growth by coalescence in coarse-grained materials, it is supplemented by grain rotation in nanocrystalline (NC) materials (Ch. 3 of \cite{koch1}). The coupled GB motion additionally plays a central role in plastic deformation of NC materials \cite{koch1}. On the other hand, the importance of junctions during microstructural evolution of polycrystalline materials is widely recognized \cite{bernacki5}. In particular for NC materials, which contain a large volume fraction of GBs (up to $30\%$) and triple junctions (up to $3\%$) \cite{meyers1}, the junctions are expected to play a central role in the coupled motion of GBs. 

Coupled GB motion has been extensively studied in two-dimensional (2D) bicrystals. Experiments with a planar GB subjected to an external stress field have shown shear deformation in the adjacent grains of a moving GB \cite{molodov1,gorkaya2}. The kinetic relations for the coupled motion were first proposed by Cahn and Taylor \cite{cahn1} who used them initially to study the evolution of an isolated  circular GB and later to non-circular GBs \cite{taylor1}. The couple motion in a bicrystal was subsequently investigated with molecular dynamics (MD) \cite{cahn2,trautt1,trautt2,bernstein1,upmanyu1}, phase field (PF) \cite{wu1,trautt2}, and level set \cite{basak1} simulations. The former two techniques have on one hand verified the phenomenon of coupled motion have otherwise provided valuable information regarding the nature of kinetic coefficients such as the geometric coupling factor. The level set simulations have studied the effect of various kinetic coefficients on the shape evolution of grains. More recently, the occurrence of coupled motion has been confirmed in the presence of junctions by MD simulations \cite{velasco1, trautt3}. It was shown that the relative rotation can sometimes get locked, thereby preventing coupled motion, but only under suitable geometric conditions (see also \cite{bernstein1, wu1}). 

In the present paper we formulate the kinetic relations for 2D coupled GB motion in the presence of junctions. We restrict our attention to a tricrystal arrangement where a grain is embedded at the center of the planar GB of a large bicrystal (motivated from \cite{trautt3}).  We obtain the governing equations for GB migration, grain rotation, as well as junction motion, all of them coupled to each other. In the absence of an external force the embedded grain shrinks towards a vanishing size. The equations are solved numerically to investigate the shape evolution of the shrinking grain for various choices of kinetic parameters and junction mobilities. The grains have been considered to be rigid. The shape accommodation process, required to avoid void formation or interpenetration at the GBs during relative rotation of the inner grain, has been controlled by allowing for diffusion along the GB.   
 
 \begin{figure}[t!]
\centering
\hspace{-2mm}
  \includegraphics[width=3.2in, height=1.6in] {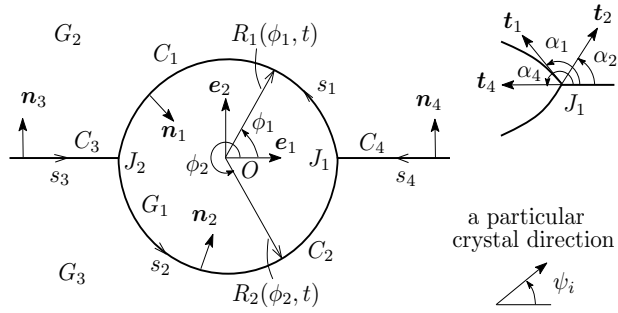}
\caption{\small{A schematic of the tricrystal. } }
\label{tricrystal_2D}
\vspace{-4mm}
\end{figure}
 
The tricrystal considered is as shown in Figure \ref{tricrystal_2D}, consisting of three grains $G_1$, $G_2$, and $G_3$, four boundaries 
$C_i$, $i=1,\ldots,4$, and two junctions $J_1$ and $J_2$. Orientation of the respective grains are $\psi_1$, $\psi_2$, and $\psi_3$, measured w.r.t. ${\boldsymbol e}_1$-axis of a fixed coordinate system with origin $O$.
The misorientation angles along $C_1$, $C_2$, and $C_{3,4}$ are defined as $\theta_1=\psi_1-\psi_2$, $\theta_2=\psi_1-\psi_3$, and $\theta_3=\psi_2-\psi_3$, respectively. The arc-length parameter for $C_i$ is denoted by $s_i$ ($i=1,\ldots, 4)$ with an increasing  direction as shown in Figure \ref{tricrystal_2D}. The normal ${\boldsymbol n}_i$ and the tangent ${\boldsymbol t}_i$ for a GB $C_i$ is also shown in the same figure, where the latter is aligned in the direction of increasing $s_i$. Let $R_i(\phi_i,t)$ be the radial distance of the GB $C_i$ from $O$ measured at an angle $\phi_i$ w.r.t. ${\boldsymbol e}_1$-axis (see Figure \ref{tricrystal_2D}), where $0\leq \phi_1\leq\pi$ and $\pi\leq\phi_2\leq2\pi$. We assume the grains to have vanishing stored energy and fixed mass density $\rho$. There is no external force applied on the tricrystal and no mass exchange with the environment. Isothermal condition is assumed throughout. The outer grains $G_1$ and $G_2$ are assumed to remain stationary, consistent with the observations made through MD simulations in \cite{trautt2}. Hence, $\dot\theta_1=\dot\theta_2=\dot\psi_1$ and $\dot\theta_3=0$, where the superposed dot represents the time derivative. We neglect rigid body translation of the grains. The mass balance at the GBs and the junctions imply (see the supplementary document for details)
\begin{equation}
\rho({\boldsymbol v}^+-{\boldsymbol v}^-)\cdot{\boldsymbol n}_i=-\frac{\partial h_i}{\partial s_i} \,\,{\rm on}\,\, C_i,\,\, \text{for}~ i=1,\ldots,4,
\label{balance_mass_2Djuncex}
\end{equation}
\begin{equation}
h_1-h_2-h_4=0 \,\,{\rm at}\,\, J_{1},~{\rm and}~ h_1-h_2+h_3=0 \,\,{\rm at}\,\, J_{2},
\label{balance_mass_junctex}
\end{equation}
 where $h_i$ is the diffusion flux along $C_i$ (positive in the direction of increasing $s_i$), ${\boldsymbol v}^+$ is the limiting value of the particle velocity as the respective boundary is approached from the side into which GB normal points, and ${\boldsymbol v}^-$otherwise. There is no summation implied with repeated indices. These equations yield $h_a = ({\rho \dot\psi_1}/{2})(R_a^2-\overline{R^2})$, for $a=1,2$, and $h_3 = h_4 = 0$, where $\overline{R^2}=\left(\int_{{C}_1}R_1^2\,dl+\int_{{C}_2}R_2^2\,dl\right)/(|{C}_1|+|{C}_2|)$ is the mean square radius of the embedded grain, $|{C}_a|$ is the length of ${C}_a$, and $dl$ is a measure of an infinitesimal arc length along the GBs.

The dissipation inequalities, in confirmation with the second law of thermodynamics, take the form \cite{basak1} (see the supplement for details)
\begin{equation}
\rho\mu({\boldsymbol v}^+-{\boldsymbol v}^-)\cdot{\boldsymbol n}_i+f_iV_i-\frac{\partial\gamma_i}{\partial\theta_i}\dot\theta_i-h_i\frac{\partial\mu}{\partial s_i}\geq 0 \,\,{\rm on}\,\, C_i,
\label{dissi_ineq_2Djuncex}
\end{equation}
\begin{equation}
{\boldsymbol F}_1\cdot{\boldsymbol q}_1\geq 0 \,\,{\rm at}\,\, J_{1},~{\rm and}~
  {\boldsymbol F}_2\cdot{\boldsymbol q}_2\geq 0 \,\,{\rm at}\,\, J_{2},
\label{dissi_ineq_junc_condex}
\end{equation}
for $i=1,\ldots,4$, where $f_i=\gamma_i\kappa_i$ is the driving force for GB migration, $\gamma_i$ is the isotropic GB energy, $\kappa_i$ is the curvature, and $V_i$ is the normal velocity for respective GBs; $\mu$ is the chemical potential of the atoms; ${\boldsymbol q}_1$ and ${\boldsymbol q}_2$ are the junction velocities and  
\begin{equation}
{\boldsymbol F}_1=\gamma_1{\boldsymbol t}_1-\gamma_2{\boldsymbol t}_2-\gamma_4{\boldsymbol t}_4, \hspace{1mm}
{\boldsymbol F}_2=-\gamma_1{\boldsymbol t}_1+\gamma_2{\boldsymbol t}_2-\gamma_3{\boldsymbol t}_3
\label{junction_force}
\end{equation}
are the corresponding driving forces at $J_1$ and $J_2$, respectively. The inequalities in \eqref{dissi_ineq_2Djuncex} are trivially satisfied for the planar GBs, and hence they remain stationary. For the curved boundaries we use the Fick's law, $h_i=-D_i\partial\mu/\partial s_i$ ($D_i$ is the diffusivity along $C_i$), to rewrite \eqref{dissi_ineq_2Djuncex} as
\begin{equation}
V_af_a+\nu_ag_a\geq 0, \,\,\text{for}\,\, a=1,2,
\label{dissipation_ineq_tricrystal1}
\end{equation}
where $\nu_a=({\boldsymbol v}_1-{\boldsymbol v}_{a+1})\cdot{\boldsymbol t}_a$ is the relative tangential velocity of the adjacent grains and
\begin{equation}
\small{
g_a=\frac{1}{{{\boldsymbol x}_a\cdot{\boldsymbol n}_a}}\left(\frac{\partial\gamma_a}{\partial\theta_a}-\frac{\dot\psi_1}{4\bar{D}_{a}}(R_a^2-\overline{R^2})^2-\rho\mu{\boldsymbol x}_a\cdot{\boldsymbol t}_a\right),}
\label{forces_velocity_tricrystal1}
\end{equation}
is the driving force for the relative tangential motion of the grains. We have used ${\boldsymbol x}_a$ for the position vector of a point on $C_1$ and $\bar{D}_a = D_a/\rho^2$. Assuming linear kinetics, the following relations are imminent (see the supplement for details)
 \begin{equation}
V_a=\small{\frac{M_a S_a}{S_a+M_a\beta_a^2}f_a-\frac{M_a \beta_a}{S_a+M_a\beta_a^2}{\boldsymbol x}_a\cdot{\boldsymbol n}_a\,\dot\psi_1, \,\,{\rm for}\,\, a=1,2,}
\label{kinetic_normal_velocity}
\end{equation}
\begin{equation}    
V_3 = V_4=0,~{\rm and}
\label{velocity_gb34}
\end{equation}
\begin{equation}
\dot\psi_1=\small{\frac{-\displaystyle\sum_{a=1}^2\displaystyle\int_{{C}_a}\left(\displaystyle\frac{M_a\beta_a}{S_a+M_a\beta_a^2}f_a\,{\boldsymbol x}_a\cdot{\boldsymbol n}_a+\frac{\partial\gamma_a}{\partial\theta_a}\right)dl}{\displaystyle\sum_{a=1}^2\displaystyle\int_{{C}_a}\left(\displaystyle\frac{({\boldsymbol x}_a\cdot{\boldsymbol n}_a)^2}{S_a+M_a\beta_a^2}-\frac{1}{2\bar{D}_a}(R_a^2-\overline{R^2})^2\right)dl},}
\label{rotation_rate1}
\end{equation}
 where $M_a$, $S_a$, and $\beta_a$, respectively, represent the mobility, the sliding coefficient, and the geometric coupling factor corresponding to ${C}_a$ (cf. \cite{basak1}). In the absence of junctions, \eqref{kinetic_normal_velocity} and \eqref{rotation_rate1} coincide with the kinetic relations obtained for a bicrystal in \cite{basak1} (see also \cite{taylor1}). To obtain the governing equations for junctions we start with \eqref{dissi_ineq_junc_condex} and propose the following linear kinetic laws \cite{fischer1} 
\begin{equation}
{\boldsymbol q}_\delta=m_\delta{\boldsymbol F}_\delta, \,\,{\rm for}\,\, \delta=1,2,
\label{junction_kinetic_relation}
\end{equation}
 where $m_\delta\geq 0$ represents the junction mobility. Assume the junctions to be non-splitting. For a finite mobility $m_1$ the velocity of $J_1$ can be written as (see the supplement for details)
\begin{equation}
{\boldsymbol q}_1 = m_1 Q_1(\cos\alpha_4{\boldsymbol e}_1+\sin\alpha_4{\boldsymbol e}_2),
\label{junction_velocity_stress_xcompo}
\end{equation}
\begin{equation}
\hspace{-0.5mm}{\rm where}\,\, Q_1= \gamma_1\cos(\alpha_1-\alpha_4)-\gamma_2\cos(\alpha_2-\alpha_4)-\gamma_4.
\label{junction_velocity_stress11}
\end{equation}
Since $\alpha_4$ is fixed, we need to compute only two junction angles $\alpha_1$ and $\alpha_2$. The following 
nonlinear algebraic equations, which ensures compatibility at the junction, are used to obtain these angles (see the supplement for details)
 \begin{equation}
{m_1} Q_1=V_a \csc(\alpha_4-\alpha_a)~{\rm with}~a=1,2.
\label{junction_compatibility_stress12}  
\end{equation}
On the other hand when $m_1\to\infty$, the velocity of $J_1$ can be obtained as
\begin{equation}
{\boldsymbol q}_1=V_a\csc(\alpha_4-\alpha_a)(\cos\alpha_4{\boldsymbol e}_1+\sin\alpha_4{\boldsymbol e}_2)
\label{junction_compatibility_nostress_mobinf2_nd1}
\end{equation}
with either $a=1$ or $2$, along with compatibility conditions 
\begin{equation}
Q_1=0 ~\text{and}~V_1\csc(\alpha_4-\alpha_1)=V_2\csc(\alpha_4-\alpha_2).
\label{junction_compatibility_stress12_mobinf1}
\end{equation}
The governing equations at $J_2$ can be obtained similarly. Time integration of \eqref{kinetic_normal_velocity}, \eqref{rotation_rate1}, and \eqref{junction_velocity_stress_xcompo} (or \eqref{junction_compatibility_nostress_mobinf2_nd1}) will give the updated position of the curved GBs, the new orientation of the embedded grain, and the updated junction positions, respectively.  

%%%%%%%%%%%%%%%%%%%%%%%
The kinetic equations are solved numerically to investigate the shape evolution of the embedded grain. We introduce non-dimensional position and time variables as $\tilde{\boldsymbol x}={\boldsymbol x}/R_0$ and $\tilde{t}=t/t_0$,
respectively, where $t_0=R_0^2/2\gamma_0M_0$ is the time taken for an isolated circular GB of radius $R_0$, with energy $\gamma_0$ and mobility $M_0$, to shrink to a point under curvature driven migration. These non-dimensional variables can be substituted in \eqref{kinetic_normal_velocity}, \eqref{rotation_rate1}, and \eqref{junction_velocity_stress_xcompo}$-$\eqref{junction_compatibility_stress12_mobinf1}, to obtain a system of non-dimensionalized equations. This naturally introduces two non-dimensional parameters $r_1=S_0/M_0$ and $r_2=M_0R_0^2/\bar{D}$ into GB kinetics, and one non-dimensional parameter $\Lambda_\delta=2R_0m_\delta/M_0$ associated with the kinetics of $J_\delta$ \cite{basak1,upmanyu3}. We write $\gamma$, $S$, and $M$ as $\gamma(\theta)=\gamma_0\tilde\gamma(\theta)$, $S(\theta)=S_0\tilde{S}(\theta)$, and $M(\theta)=M_0\tilde{M}(\theta)$ \cite{basak1}. We restrict our simulations to constant mobility and sliding coefficient, i.e. we take $\tilde{M}=\tilde{S}=1$. We also assume $\Lambda_1 = \Lambda_2 = \Lambda$. The value of the dimensionless parameters are taken as $r_1=0.01$, $r_2=10^3({\overline{R}}(0)/{\overline{R}}(\tilde{t}))^{3/2}$, and $\infty\leq\Lambda\leq 1$ \cite{upmanyu3,basak1}. The time-dependent term in $r_2$ ensures that with decreasing grain size GB diffusivity increases \cite{chen1}. GB energy is considered to be isotropic with a form derived from a disclination model of the GB and hence valid for large misorientation angles \cite{nazarov1}. The expression for coupling factor $\beta$ is taken from Cahn et al. \cite{cahn2}. The formulas for both $\tilde{\gamma} (\theta)$ and $\beta(\theta)$ are provided in the supplement. All the parameters have been taken for fcc crystals.

Our simulation methodology is based on the finite difference scheme proposed by Fischer et al. \cite{fischer1}. We initially discretize the curved GBs with $100$ and the planar GBs with $25$ grid points. To avoid mesh points coming very close to each other or moving far away after time integration, we re-mesh the GBs after every iteration so as to maintain accuracy and stability of the numerical calculations. All the computations are done in a domain of size $[-0.6,0.6]\times [-0.6,0.6]$, with time steps as $10^{-5}$ and $10^{-4}$ for the case of GB migration and coupled GB motion, respectively.  In all the simulations initial radius of $G_1$ is taken to be $\tilde{R}_a(0)=0.4$ and all the GBs are assumed to be $[001]$ tilt boundaries. As a sign convention, if any of the misorientation angles turns out to be negative, we add $90^\circ$ to it to obtain an equivalent misorientation angle in the range $0\leq \theta_i< 90^\circ$ recalling that the considered crystals posses a four-fold symmetry \cite{trautt3}. The initial orientation of the grains are taken as $\psi_1 =14^\circ$, $\psi_2=0^\circ$, and $\psi_3 = 60^\circ$. The initial misorientations are therefore $\theta_1=14^\circ$, $\theta_2=44^\circ$, and $\theta_3=30^\circ$. During the coupled motion only $\psi_1$ (and hence $\theta_1$ and $\theta_2$) is allowed to changed while others are kept constant.

\begin{figure}[t!]
\centering
%\hspace{-4mm}
  \includegraphics[width=3.2in, height=1.6in] {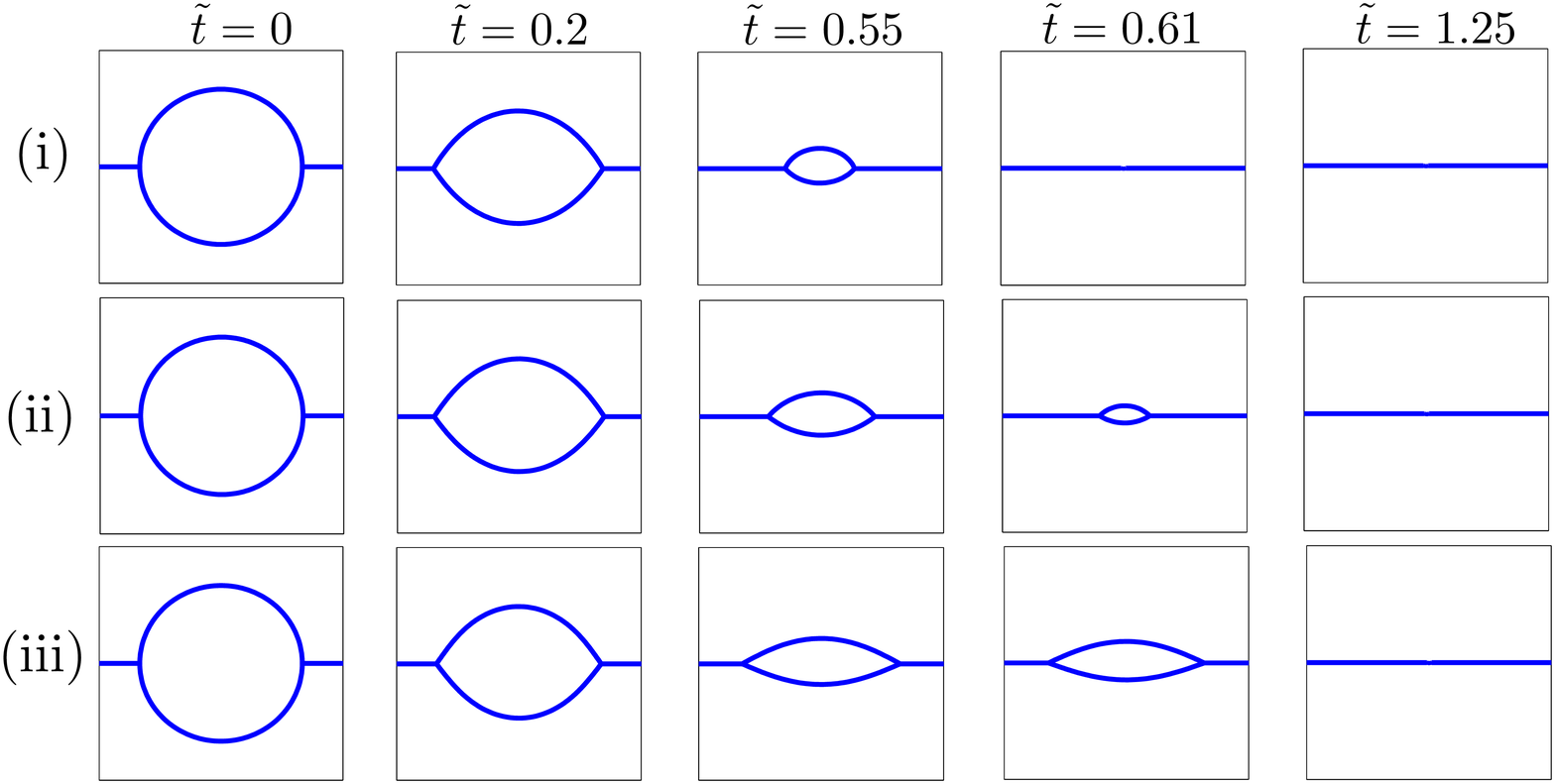}
	%	\vspace{-4mm}
 \caption{\small{Shape evolution under GB migration when $\psi_1 =14^\circ$, $\psi_2=0^\circ$, and $\psi_3 = 60^\circ$. Rows (i) to (iii) correspond to $\Lambda\to\infty$, $\Lambda=20$, and $\Lambda=1$, respectively. }}
\label{normal_motion_Lambda_compare}
\end{figure}
\begin{figure}[t!]
\centering
\hspace{-4mm}
\includegraphics[width=3.2in, height=1.6in] {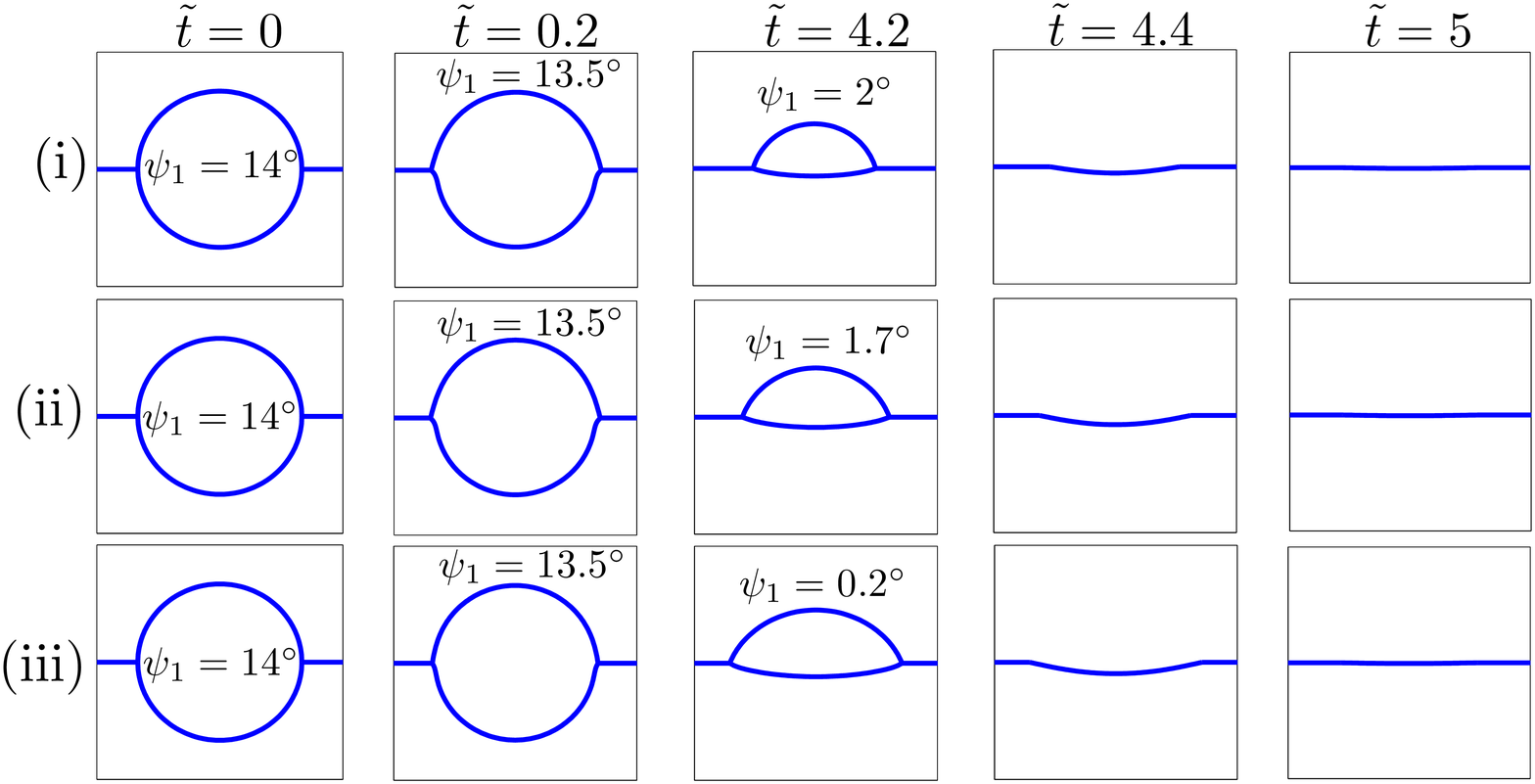}
\caption{\small{Shape evolution under fully coupled GB motion when initial $\psi_1 =14^\circ$, and $\psi_2=0^\circ$ and $\psi_3 = 60^\circ$. Rows (i) to (iii) correspond to $\Lambda\to\infty$, $\Lambda=20$, and $\Lambda=1$, respectively.}}
\vspace{-4mm}
\label{coupled_Sbeta_nonzero_Lambda_infty_20_1}
\end{figure}

\begin{figure*}[t!]
\centering
  \includegraphics[width=6in, height=3in] {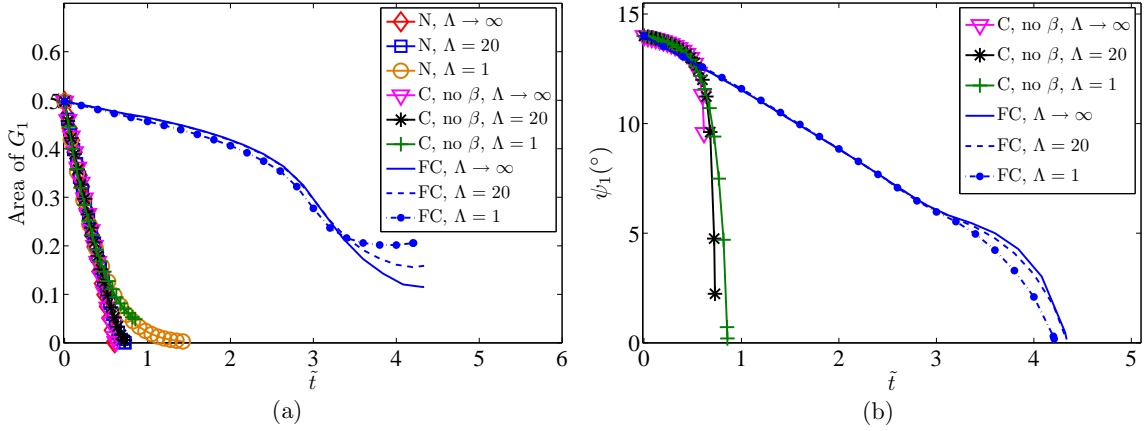}
\vspace{-10mm}
    \caption{\small{(a) Area and (b) orientation evolution of the embedded grain under normal and coupled GB motion.  Abbreviations: N - GB migration, C - coupled GB motion in the absence of $\beta_1$ and $\beta_2$, and FC - fully coupled GB motion.}}
		%\vspace{-1mm}
		\label{area_orientation_embedded_grain}
\end{figure*}

{\em GB migration:} With $\beta \to 0$ and ${S}\to 0$ the kinetic relations \eqref{kinetic_normal_velocity} and \eqref{rotation_rate1} are reduced to $\tilde{V}_a=\tilde{M}\tilde\gamma\tilde\kappa/2$ and ${\dot \psi}_1=0$, respectively.
Figure \ref{normal_motion_Lambda_compare} shows the evolution of the embedded grain under these assumptions with both finite and infinite junction mobility. The junction angles start evolving soon after the evolution starts and the embedded grain attains a lens shape. A finite junction mobility drags the GB motion and retards the shrinking rate of the embedded grain. The drag effect increases as $\Lambda$ decreases and the curved GBs become increasingly flatter before shrinking (see also Figure \ref{area_orientation_embedded_grain}). However, the junction velocities become comparable with those of the GBs when $\Lambda>>1$, which reduces the drag on the GBs. The area evolution then becomes nearly linear and the deviation from linearity increases as $\Lambda$ decreases. The effect of finite junction mobility has been widely noticed to have a significant influence on GB dynamics (see for e.g. \cite{upmanyu3,zollner1}). The drag effects at the junctions are due to frequent dislocation reactions and changes in point defect density in their vicinity (Ch. 3 in \cite{koch1}). 

{\em Coupled GB motion:} Depending on the operating conditions, some of the kinetic parameters may be more active than the others. For example, at temperatures near the melting point, viscous GB sliding dominates over geometric coupling, whereas at relatively lower temperatures, sliding becomes much less active than geometric coupling \cite{cahn2}.  We demonstrate the effect of kinetic coefficients on the shape evolution by considering several cases below. 

(i) {\em Fully coupled:} When both sliding and geometric coupling are active, the grain shrinkage becomes much slower than with GB migration alone, as shown in Figures \ref{coupled_Sbeta_nonzero_Lambda_infty_20_1} and \ref{area_orientation_embedded_grain}. However, the combined effect of the GB energy and the kinetic coefficients is such that the lower GB shrinks faster than the upper one. Also note that the vanishing of $\psi_1$ (and hence $\theta_1$) results into a bicrystal with a depression on the GB, which ultimately disappears to yield a perfectly planar GB. Moreover, the finite junction mobility not only drags the GB motion, but also slows down the grain rotation, as can be seen in Figure \ref{area_orientation_embedded_grain}(b).

%\begin{figure}[t!]
%\centering
  %\includegraphics[width=3.5in, height=1.8in] {./figures/area_plot_normal_coupled.eps}
		%\includegraphics[width=3.5in, height=1.8in] {./figures/orientation_angle1_plot_normal_coupled.eps}
    %\caption{\small{(a) Area and (b) orientation evolution of the embedded grain under normal and coupled GB motion.  Abbreviations: N - GB migration, C - coupled GB motion in the absence of $\beta_1$ and $\beta_2$, and FC - fully coupled GB motion.}}
		%\vspace{-4mm}
		%\label{area_orientation_embedded_grain}
%\end{figure}

(ii) {\em No geometric coupling:} In the absence of $\beta$, the non-dimensional equation for normal velocity  reduces down to $\tilde{V}_a= \tilde{M}_a\tilde\gamma_a\tilde\kappa_a/2$, which is same as the evolution equation for GB migration, except for the fact that $\tilde\gamma_a$ is now evolving with time (due to evolving misorientation). 
Figure \ref{area_orientation_embedded_grain}(a) shows that the area evolution is now slightly slower than the case of GB migration. Orientation $\psi_1$ evolves very slowly for most of the time except towards the end.  The shape evolution of the curved GBs is nearly identical to the ones shown in Figure \ref{normal_motion_Lambda_compare} for respective junction mobilities. When $\Lambda\to\infty$ and $\Lambda=20$, the grain shrinks before $\psi_1$ could vanish. However when $\Lambda=1$, $\psi_1$ vanishes before the area,  leaving a bicrystal with depression on the planar GB, which eventually vanishes. 

(iii) {\em No sliding:} When $S \to 0$, \eqref{kinetic_normal_velocity} implies that the GB shape $R_a(\phi_a,t)$ will remain self-similar for all times as long as $\beta$ is isotropic \cite{taylor1, basak1}.
 For example, if $G_1$ is initially a circle, then it should remain so for all times during the evolution. Obviously with such a restriction, compatibility equations \eqref{junction_compatibility_stress12} or \eqref{junction_compatibility_stress12_mobinf1} will have solutions only for very special initial geometries of $C_1$ and $C_2$. 

Finally, take orientation $\psi_3$ to be $28^\circ$ keeping initial $\psi_1$ and $\psi_2$ same as above. The initial misorientations are therefore $\theta_1=14^\circ$, $\theta_2 = 76^\circ$, and $\theta_3=62^\circ$. As a result the curved GBs are symmetrically equivalent with $\beta_1=-\beta_2$. Since the embedded grain is initially symmetric about ${\boldsymbol e}_1$-axis, the first term in the numerator of \eqref{rotation_rate1} disappears. However, for the GB energy considered here, the second term in the numerator will lead to non-zero rotation of $G_1$. On the other hand, if the energy is symmetric about $\theta=45^\circ$ (as is the case with the energy given in Figure 4 of \cite{gjostein1}), the rotation of $G_1$ will vanish and it will shrink purely by migration of $C_1$ and $C_2$. This phenomenon of rotation getting  locked has been observed in MD \cite{trautt2,bernstein1} and PF simulations \cite{wu1} when $C_1$ and $C_2$ are symmetrically equivalent. 

In summary, we have presented an analytical framework to study coupled GB motion in a tricrystal. We have proposed a system of kinetic relations which govern GB motion, grain rotation, as well as junction motion. While investigating the role of junction we have observed that decreasing junction mobility can significantly decelerate the evolution of shape, area, and orientation of the embedded grain. The effect of various kinetic coefficients has also been emphasized.

 %\begin{figure}[t!]
%\centering
%\subfigure[ ]{
  %\includegraphics[width=2.8in, height=1.4in] {./figures/area_plot_normal_coupled.eps}\quad
		%\label{area_plot_normal_coupled}}\nonumber
		%\vspace{-6mm}
    %\subfigure[ ]{
    %\includegraphics[width=2.8in, height=1.4in] {./figures/orientation_angle1_plot_normal_coupled.eps}
    %\label{orientation_angle1_plot_normal_coupled}}
		%\vspace{-5mm}
%\caption{\small{(a) Area and (b) orientation evolution of the embedded grain under normal and coupled GB motion.  Abbreviations: N - normal GB motion, C - coupled GB motion in absence of $\beta_1$ and $\beta_2$, and FC - fully coupled GB motion.}}
%\vspace{-4mm}
%\label{area_orientation_embedded_grain}
%\end{figure}

%%%%%%%%%%%%%%%%%%%%%%%%%%%%%%%%%%%%%%%%%%%%%%%%%
%\vspace{4mm}
\small{The authors gratefully acknowledge Professor Jiri Svoboda for helpful discussions.}
\begingroup
\renewcommand{\section}[2]{}

%%%%%%%%%%%%%%%%%%%%%%%%%%%%%%%%%%%%%%%%%%%%
\begin{thebibliography}{99}
%%%%%%%%%%%%%%%%%%%%%%%%%%%%%%%%%%%%%%
\small{
\setlength{\parskip}{-0.1cm}
\bibitem{cahn1} 
\newblock{J.W. Cahn, J.E. Taylor, Acta Mater. 52 (2004) 4887.}
%\newblock {\em Mechanical Properties of Nanocrystalline Materials}.
%%%%%%%%%%%%%%%%%%%%%%%%%%%%%%%%%%
\bibitem{koch1} 
\newblock{C.C. Koch, I.A. Ovid$'$ko, S. Seal, S. Veprek, Structural Nanocrystalline Materials: Fundamentals and Applications, Cambridge University Press, New York, 2007.}
%%%%%%%%%%
\bibitem{bernacki5} 
\newblock{M. Bernacki, Y. Chastel, T. Coupez, R.E. Lo\'ge, Scr. Mater. 58 (2008) 1129.}
%TITLE={Level Set Framework for the Numerical Modelling of Primary Recrystallization in Polycrystalline Materials},
%%%%%%%%%%%%%%%%%%%%%%%%%%%%%%%%%%
\bibitem{meyers1} 
\newblock{M.A. Meyers, A. Mishra, J.D. Benson, Prog. Mater. Sci. 51 (2006) 427.}
%TITLE={Mechanical Properties of Nanocrystalline Materials},
%%%%%%%%%%%%%%%%%%%%%%%%%%%%%%%%
\bibitem{gorkaya2} 
\newblock{T. Gorkaya, T. Burlet, D.A. Molodov, G. Gottstein, Scr. Mater. 63 (2010) 633.}
%TITLE={Experimental Method for True in Situ Measurements of Shear-coupled Grain Boundary Migration},
%%%%%%%%%%%%%%%%%%%%%%%%%%%%%%%%%%
\bibitem{molodov1} 
\newblock{D.A. Molodov, V.A. Ivanov, G. Gottstein, Acta Mater. 55 (2007) 1843.}
%TITLE={Low Angle Tilt Boundary Migration Coupled to Shear Deformation},
%%%%%%%%%%%%%%%%%%%%%%%%%%%%%%%%%%
\bibitem{taylor1} 
\newblock{J.E. Taylor, J.W. Cahn, Interfaces Free Bound. 9 (2007) 493.}
%TITLE={Shape Accommodation of a Rotating embedded crystal via a New Variational Formulation},

%%%%%%%%%%%%%%%%%%%%%%%%%%%%%%%%%%
\bibitem{bernstein1} 
\newblock{N. Bernstein, Acta Mater. 56 (2008) 1106.}
%TITLE={The Influence of Geometry on Grain Boundary Motion and Rotation},
%%%%%%%%%%%%%%%%%%%%%%%%%%%%%%%%%%
\bibitem{trautt1} 
\newblock{Z.T. Trautt, Y. Mishin, Acta Mater. 60 (2012) 2407.}
%TITLE={Grain Boundary Migration and Grain Rotation Studied by Molecular Dynamics},
%%%%%%%%%%%%%%%%%%%%%%%%%%%%%%%%%%
\bibitem{trautt2} 
\newblock{Z.T. Trautt, A. Adland, A. Karma, Y. Mishin, Acta Mater. 60 (2012) 6528.}
%TITLE={Coupled Motion of Asymmetrical Tilt Grain Boundaries: Molecular Dynamics and Phase Field Crystal Simulations},
%%%%%%%%%%%%%%%%%%%%%%%%%%%%%%%%%%
\bibitem{cahn2} 
\newblock{J.W. Cahn, Y. Mishin, A. Suzuki, Acta Mater. 54 (2006) 4953.}
%%%%%%%%%%%%%%%%%%%%%%%%%%%%%%%%%%
\bibitem{upmanyu1} 
\newblock{M. Upmanyu, D.J. Srolovitz, A.E. Lobkovsky, J.A. Warren, W.C. Carter, Acta Mater. 54 (2006) 1707.}
%TITLE={Simultaneous Grain Boundary Migration and Grain Rotation},
%%%%%%%%%%%%%%%%%%%%%%%%%%%%%%%%%%
\bibitem{wu1} 
\newblock{K.A. Wu, P.W. Voorhees, Acta Mater. 60 (2012) 407.}
%TITLE={Phase Field Crystal Simulations of Nanocrystalline Grain Growth in Two Dimensions},
%%%%%%%%%%%%%%%%%%%%%%%%%%%%%%%%%%%%
\bibitem{basak1} 
\newblock{A. Basak, A. Gupta, Modelling Simul. Mater. Sci. Eng. (accepted for publication).\\
Preprint available at: http://home.iitk.ac.in/${\sim}$ag/papers/BasakGupta14r.pdf}
%TITLE{A Two-dimensional Study of Coupled Grain Boundary Motion Using Level Set Method},
%%%%%%%%%%%%%%%%%%%%%%%%%%%%%%%%%%
\bibitem{trautt3} 
\newblock{Z.T. Trautt, Y. Mishin, Acta Mater. 65 (2014) 19.}
%TITLE={Capillary-driven Grain Boundary Motion and Grain Rotation in a Tricrystal: {A} Molecular Dynamics Study},
%%%%%%%%%%%%%%%%%%%%%%%%%%%%%%%%%%
\bibitem{fischer1} 
\newblock{F.D. Fischer, J. Svoboda, K. Hackl, Acta Mater. 60 (2012) 4704.}
%TITLE={Modelling the Kinetics of a Triple Junction},
%%%%%%%%%%%%%%%%%%%%%%%%%%%%%%%%%%
\bibitem{upmanyu3} 
\newblock{M. Upmanyu, D.J. Srolovitz, L.S. Shvindlerman, G. Gottstein, Acta Mater. 50 (2002) 1405.}
%TITLE={Molecular Dynamics Simulation of Triple Junction Migration},
%%%%%%%%%%%%%%%%%%%%%%%%%%%%%%%%%%
\bibitem{chen1} 
\newblock{Y. Chen, C.A. Schuh, J. Appl. Phys. 101 (2007) (063524) 1.}
%TITLE={Geometric Considerations for Diffusion in Polycrystalline Solids},
%%%%%%%%%%%%%%%%%%%%%%%%%%%%%%%%%%
\bibitem{nazarov1} 
\newblock{A.A. Nazarov, O.A. Shenderova, D.W. Brenner, Mater. Sci. Eng. A. 281 (2000) 148.}
%TITLE={On the Disclination-Structural Unit Model of Grain Boundaries},
%%%%%%%%%%%%%%%%%%%%%%%%%%%%%%%
\bibitem{velasco1} 
\newblock{M. Velasco, H.V. Swygenhoven, C. Brandl, Scr. Mater. 65 (2011) 151.}
%TITLE={Coupled Grain Boundary Motion in a Nanocrystalline Grain Boundary Network},
%%%%%%%%%%%%%%%%%%%%%%%
\bibitem{zollner1} 
\newblock{D. Z\"ollner, Scr. Mater. 67 (2012) 41.}
%TITLE={Grain Microstructure Evolution in Two-dimensional Polycrystals Under Limited Junction Mobility},
%%%%%%%%%%%%%%%%%%%%%%%%%%%%%%%%%%
\bibitem{gjostein1} 
\newblock{N.A. Gjostein, F.N. Rhines, Acta Metall. 7 (1959) 319.}
%TITLE={ABSOLUTE INTERFACIAL ENERGIES OF [OO1] TILT AND TWIST GRAIN BOUNDARIES IN COPPER},

%%%%%%%%%%%%%%%%%%%%%%%%%%%%%%%%%%
%\bibitem{shih1} 
%\newblock{K.K. Shih, J.C.M. Li, Surf. Sci. 50 (1975) 109.}
%TITLE={Energy of Grain Boundaries Between Cusp Misorientations},
%%%%%%%%%%%%%%%%%%%%%%%%%%%
%\bibitem{novikov1} 
%\newblock{V. Y. Novikov, Scr. Mater. 52 (2005) 857.}
%TITLE={On the Influence of Triple Junctions on Grain Growth Kinetics and Microstructure Evolution in 2D Polycrystals},
%%%%%%%%%%%%%%%%%%%%%%%%%%%%%%
%\bibitem{chokshi1} 
%\newblock{A.H. Chokshi, Scr. Mater. 59 (2008) 726.}
%%TITLE={Triple Junction Limited Grain Growth in Nanomaterials},

%\bibliography{junction_2D}
%\bibliographystyle{nar}
}
\endgroup
%\end{thebibliography}

%%%%%%%%%%%%%%%%%%%%%%%%%%%%%%%%%%%%%%%%%%%%%%%%%
\end{document}